\documentclass[pra,12pt,floatfix]{revtex4}
\usepackage{graphicx}
\usepackage{amsmath}

\begin{document}

\title{Comparative study of metal cluster fission\\ in
Hartree-Fock and LDA}

\author{Andrey  Lyalin}
\email[Email address: ]{lyalin@rpro.ioffe.rssi.ru}
\affiliation{Institute of Physics, St Petersburg State University,
 Ulianovskaja 1, St Petersburg, Petrodvorez, Russia 198504}

\author{Andrey Solov'yov}
\altaffiliation[Permanent address: ]{A. F. Ioffe Physical-Technical
Institute of the Russian Academy of Sciences, Polytechnicheskaya 26,
St. Petersburg, Russia 194021}
\email[Email address: ]{solovyov@rpro.ioffe.rssi.ru}

\author{Walter Greiner}
\affiliation{Institut f\"{u}r Theoretische Physik der Universit\"{a}t
Frankfurt am Main, Robert-Mayer 8-10, Frankfurt am Main, Germany 60054}

\begin{abstract}

Fission of doubly charged metal clusters 
is studied using the open-shell two-center deformed jellium Hartree-Fock model
and Local Density Approximation.
Results of calculations of the electronic structure and fission barriers
for the symmetric and asymmetric channels associated with the
following processes
$Na_{10}^{2+} \rightarrow Na_{7}^{+} + Na_{3}^{+}$,
$Na_{18}^{2+} \rightarrow Na_{15}^{+} + Na_{3}^{+}$ and
$Na_{18}^{2+} \rightarrow 2 Na_{9}^{+}$
are presented. The role of the exact exchange and
many-body correlation effects in metal clusters fission
is analysed.
It is demonstrated that the 
influence of many-electron correlation
effects on the height of the fission barrier is more profound
if the barrier arises nearby or beyond
the scission point.
The importance of cluster deformation effects in the fission process
is elucidated with the use of the overlapping-spheroids shape parametrization
allowing one an independent variation of deformations in the parent
and daughter clusters.

\end{abstract}



\maketitle

\section{Introduction}

Fission of charged atomic clusters occurs when repulsive Coulomb forces,
arising due to the excessive  charge, overcome the electronic binding energy
of the cluster \cite{Sattler81,Naher97,Yannouleas99}.
This mechanism of the cluster
fission is in a great deal similar to the nuclear fission phenomena.
Experimentally, multiply charged metal clusters
can be observed in the mass spectra
when their size exceeds the critical size of stability, which depends
on the metal species and cluster charge
\cite{Brechignac89,Brechignac90,Brechignac91,Brechignac94,Martin84,Martin92}.
For clusters above critical size, simple evaporation of neutral
species is the dominant fragmentation channel,
while below the critical size, fission into two charged
fragments is more probable. At a low temperature fissioning
clusters can be metastable above a certain size, because of the
existence of a fission barrier.

Initially, theoretical studies of cluster stability were based
on pure energetic criteria that only involved the energies
of the initial and the final states \cite{Tomanek83,Iniguez86,Rao87}.
Later, a simple one center Liquid Drop Model (LDM),
which was initially suggested by Lord Rayleigh in 1882 \cite{Rayleigh1882}
and later widely used in nuclear physics,
was adapted to charged metal clusters \cite{Saunders92}.
In this model, one introduces the "fissility parameter",
$X=E^{shere}_{Coul}/2E^{sphere}_{Surf}$, which is proportional to the ratio
of the Coulomb to surface energy of charged spherical liquid drop
\cite{Rayleigh1882}. 
The fissility parameter distinguishes
the situations when cluster is unstable, metastable, or stable.
The investigation of the Rayleigh instabilities in multiply charged
sodium clusters has been done in \cite{Guet01a}, where reasonable agreement
with experimental data was found.

In spite of the fact that the simple LDM qualitatively describes the
fission process it fails to reproduce experimental data in full detail.
This happens because the LDM does not take into account
shell effects. It has been shown that the shell effects are important in
nuclear fission \cite{Eisemberg_and_Greiner} and even more important in 
fission of metal clusters \cite{Brechignac94}. One can
describe the shell effects in metal clusters
using the Shell Correction Method (SCM), originally developed in nuclear
physics \cite{Strutinsky67,Strutinsky68,Eisemberg_and_Greiner}.  This method was adapted
for metal clusters in
\cite{Nakamura90,Yannouleas93a,Reimann93,Koizumi93,Yannouleas95,Frauendorf96}.

The asymmetric two-center-oscillator shell model
(ATCOSM), introduced in \cite{Maruhn72} for nuclear fission
is quite successful in prediction of the fission barriers.
This model was also applied for the description
of metal cluster fission  \cite{Yannouleas95a}.
Although ATCOSM method uses single electron model potential it has
the significant advantage in comparison to other models allowing
one simple shape parametrization and an independent variation of
deformations in the parent and daughter clusters \cite{Yannouleas95a}.

The microscopic description of energetics and
dynamics of metal cluster fission process based on molecular dynamic
(MD) simulations has been performed in
\cite{Barnett91,Brechignac94a,Montag95a,Guet01} using the
local-spin-density-functional method. This method is,
however, strongly restricted by the  cluster size, because of computational
difficulties, and thus is usually applied to the small metal clusters with the
number of atoms $N \le 20$.

Fission process of metal clusters can be also simulated on the
basis of the jellium model, which does not take into account the detailed
ionic structure of the cluster core.
Jellium model considers the electrons in the usual quantum mechanical way,
but approximates the cluster core potential by the potential of the
homogeneous positively charged background and, therefore, is better
applicable for lager cluster sizes (see e.g. \cite{Ekard-book} for review).
Most of the electronic structure calculations of the jellium metal clusters
have been performed using self-consistent Kohn-Sham Local
Density Approximation (LDA) \cite{Kohn-Sham}.
The LDA jellium calculations for metal cluster fission
can be grouped \cite{Yannouleas95a} into two categories
according to the fragment shape parametrization, namely, the
two-intersected-spheres jellium \cite{Saito87,Saito88,Knospe93} and
variable-necking-in parametrization \cite{Garcias94,Garcias95,Koizumi95}.
It has been shown that the cluster shape parametrization must be
flexible enough to account for the majority of  effects generated
by the shell structure of the parent and daughter clusters, which in general
have not spherical but deformed shapes \cite{Yannouleas95a}.

The important feature of the LDA method consists in the fact that
it takes into account many-electron correlations via the
phenomenological exchange-correlation potential
(see e.g. \cite{Parr-book,Fulde-book} for review).
However, so far, there has not been found the unique potential,
universally applicable for different systems and conditions.
As a result there is a "zoo of potentials" \cite{zoo} valid for
special cases. These potentials, of course,
do exist in principle as unique quantities but are not actually understood,
so they cannot serve as a satisfactory basis for achieving a
physical interpretation.

Alternatively, one can develop direct  {\it ab initio} methods
for the description of electronic properties of metal clusters. It can be
achieved by using the Hartree-Fock (HF) approximation and by the construction
on its basis the systematic many-body theories such as the random phase
approximation with exchange \cite{Amusia90}, many-body perturbation
theory or the Dyson equation method
\cite{Chernysh88,Grib90,Ivanov96}.  Based on fundamental physical
principles these models can be refined by extending the quality of the
approximations, while the physical meaning of the effects included are
clearly demonstrated and thus give more accurate characteristics of
metal clusters than LDA.

Originally, the Hartree-Fock model for the metal cluster electron structure
has been worked out in the framework of spherically symmetric jellium
approximation in \cite{GJ92,IIKZ93}. It is valid for the clusters with closed
electronic shells that correspond to magic numbers (8, 20, 34, 40,...).
On the basis of the Hartree-Fock approach
the dynamic jellium model has been proposed \cite{GSG99,GISG00}.
This model treats simultaneously the vibrational modes of the ionic jellium
background, the quantized electron motion and the interaction
between the electronic and the ionic subsystems.
In particular, the dynamic jellium model allows to describe the widths
of electron excitations in metal clusters beyond the adiabatic
approximation.
The open-shell two-center jellium Hartree-Fock approximation valid
for metal clusters with arbitrary number of the valence
electrons has been developed in \cite{Lyalin00,Lyalin01a}.
The two-center jellium HF method treats the quantized electron motion in the
field of the spheroidal ionic jellium background
in the spheroidal coordinates. This method has been 
generalized and adopted to study of the metal clusters fission process
in our recent work \cite{Lyalin01b}, where barrier
for the symmetric fission cannel 
$Na_{18}^{2+} \rightarrow 2 Na_{9}^{+}$
was calculated.

In the present work we investigate
the role of the exchange and correlation effects in
metal cluster fission process on the basis of both the
Hartree-Fock and LDA methods.
Both symmetric and asymmetric fission
channels for the $Na_{10}^{2+}$ and $Na_{18}^{2+}$ parent clusters
are considered.
Comparison of results of the two approaches allows us to illustrate
the importance of the exchange component of the many-electron 
interaction in the fission process and make important conclusion
about the relative role of the two different channels of the reaction.

The atomic system of units, $|e|=m_e=\hbar=1$, has been used
throughout the paper, unless other units not-indicated.

\section{Theoretical methods}

\subsection{Two-center jellium model and cluster shape parametrization}

According to the main postulate of the jellium model,
the electron motion in a metallic cluster takes place in the field of
the uniform positive charge distribution of the ionic background.
For the parametrization of the ionic background during the fission process we
consider the model in which the initial parent cluster having the form of the
ellipsoid of revolution (spheroid) splits into two independently deformed
spheroids of smaller size \cite{Lyalin01b}.
The two principal diameters $a_{k}$ and $b_{k}$
of the spheroids can be expressed via the deformation parameter
$\delta_{k}$ as
\begin{equation}
a_{k} = \left( \frac{2+\delta_{k}}{2-\delta_{k}}\right)^{2/3}R_{k}, \ \ \ \
b_{k} = \left( \frac{2-\delta_{k}}{2+\delta_{k}}\right)^{1/3}R_{k}.
\label{eq:axes}
\end{equation}

\noindent Here partial indexes $k=0,1,2$ correspond to the parent 
cluster ($k=0$) and the two daughter fragments ($k=1,2$),
$R_{k}$ ($k=0,1,2$) are the
radii of the corresponding undeformed spherical cluster.
The deformation parameters $\delta_{k}$ characterize the families
of the prolate ($\delta_{k} > 0$) and the oblate
($\delta_{k} < 0$) spheroids of equal volume $V_{{k}}=4\pi a_{k}
b_{k}^{2}/3=4\pi R_{k}^3/3$.

The radii of the parent and the resulting non overlapping daughter fragments
are equal to $R_{k}=r_s N_{k}^{1/3}$, where $N_{k}$ is the number of atoms
in the $k$th cluster,
and $r_s$ is the Wigner-Seitz radius. For sodium clusters, $r_s=4.0$,
which corresponds to the density of the bulk sodium. For overlapping region the
radii $R_{1}(d)$ and $R_{2}(d)$ are functions of the distance $d$
between the centers of mass of the two fragments.  They are so determined
that the total volume inside the two spheroids equals the volume of the parent
cluster $4\pi R_{0}^3/3$.

The ions charge density $\rho({\bf r})$ is kept
uniform including the overlapping-spheroids region,
\begin{equation}
\label{rho}
\rho({\bf r})=
\begin{cases}
\rho_c, & {(x^2+y^2)}/{b_{1}^2} + {(z+d/2)^2}/{a_{1}^2} \leq 1 \\
\rho_c, & {(x^2+y^2)}/{b_{2}^2} + {(z-d/2)^2}/{a_{2}^2} \leq 1 \\
0,      & \text{otherwise}.
\end{cases}
\end{equation}

\noindent Here $\rho_c=Z_{{0}}/V_{{0}}$ is the ionic charge density
inside the cluster and $Z_{{0}}$ is the total charge
of the ionic core.

The electrostatic potential $U_{core}({\bf r})$ of the ionic background can be
determined from the solution of the corresponding Poisson equation:
\begin{equation}
\Delta U_{core}({\bf r}) = - 4\pi \rho({\bf r}).
\label{peq}
\end{equation}

\subsection{Hartree-Fock and LDA formalism}

The Hartree-Fock equations can be written out explicitly in the form
(see, e.g., \cite{Lindgren}):
\begin{equation}
\left( - \Delta/2 + U_{core} + U_{HF}\right) \mid a > \, = \,
\varepsilon_a \mid a >.
\label{HF}
\end{equation}

The first term here represents the kinetic energy of electron $a$,
and $U_{core}$ its attraction to the cluster core. The Hartree-Fock
potential $U_{HF}$ represents the average Coulomb interaction
of electron $a$ with the other electrons in the cluster, including
the non-local exchange interaction,
and $\varepsilon_a$ describes the single electron energy.

According to the Density Functional Theory (DFT),
the ground state energy is a minimum for the exact density
of the functional of the density of the system \cite{Hohenberg-Kohn}.
A self-consistent method for calculation
of the electronic states of many-electron systems was proposed by
Kohn and Sham \cite{Kohn-Sham}. This method leads to the
Kohn-Sham LDA self consistent equations, which are similar to
the Hartree equations:   
\begin{equation}
\left( - \Delta/2 + U_{core} +  U_{H} + V_{xc}\right) \mid a > \, = \,
\varepsilon_a \mid a >.
\label{LDA}
\end{equation}

\noindent Here $U_{H}$ is the Hartree potential, which represents the direct
Coulomb interaction of electron $a$ with the other electrons in the cluster,
but does not take into account the non-local exchange effects, while
$V_{xc}$ is the phenomenological density dependent local
exchange-correlation potential. 
In the present work we use the Gunnarsson and Lundqvist
model \cite{Gunnarsson}
for the LDA electron exchange-correlation energy density $\epsilon_{xc}$,
which reads as
\begin{equation}
\epsilon_{xc}(\rho_{el}({\bf r})) =
-\frac{3}{4}\left( \frac{9}{4 \pi^2} \right)^{1/3}
\frac{1}{r_s({\bf r})} -
0.0333\ G \left( r_s({\bf r})/11.4 \right).
\label{GunLun}
\end{equation}

\noindent Here
$r_s({\bf r})=(3/4\pi\rho_{el}({\bf r}))^{1/3}$ is a
{\it local} Wigner-Seitz radius, while $\rho_{el}({\bf r})$
is the electron density in the cluster, and the function $G(x)$
is defined by following relation:
\begin{equation}
G(x)= (1+x^3)\ln \left(1+\frac{1}{x} \right) -
x^2 +  \frac{x}{2} - \frac{1}{3}.
\label{Gx}
\end{equation}

The exchange-correlation energy density $\epsilon_{xc}(\rho_{el}({\bf r}))$,
defines the LDA exchange-correlation potential $V_{xc}(\rho_{el}({\bf r}))$ as 
\begin{eqnarray}
V_{xc}(\rho_{el}({\bf r})) &=& \frac{\delta\left[
\rho_{el}({\bf r})\epsilon_{xc}(\rho_{el}({\bf r}))\right]}
{\delta\rho_{el}({\bf r})} = \\ \nonumber
&& -\left( \frac{9}{4 \pi^2} \right)^{1/3} \frac{1}{r_s({\bf r})} -
0.0333\ \ln\left(1+ \frac{11.4}{r_s({\bf r})} \right).
\label{LDA-pot}
\end{eqnarray}

The Hartree-Fock (\ref{HF}) and LDA (\ref{LDA}) equations
have been solved in the system of the prolate spheroidal coordinates
as a system of coupled two-dimensional second order partial differential 
equations. The partial differential equations have been discretized on a
two-dimensional grid and the resulting system of linear equations has been
solved numerically by the successive overrelaxation method \cite{SOR}.  
This technique is different from that we have used in our previous works
\cite{Lyalin00,Lyalin01a,Lyalin01b}, where the expansion of wave
functions and potentials over spheroidal harmonics in the prolate and oblate
spheroidal coordinates in one dimension and a numerical expansion
for the second dimension have been carried out. The third dimension,
the azimuthal angle, has been treated analytically in both methods.
Our calculations show that the partial-expansion method is effective for 
slightly deformed systems, for which only few terms in the expansion 
are necessary to take into account in order to achieve
sufficient accuracy.
However, for strongly deformed systems, or process like fission,
the direct two-dimensional
integration is more efficient.

The important characteristic of the cluster,
which defines its stability is the total energy $E_{tot}$.
The total energy of the cluster is
equal to the sum of the electrostatic energy of the ionic core
$E_{core}$  and the energy of the valence electrons
$E_{el}$:
\begin{equation}
    E_{tot}=E_{core} + E_{el}.
\label{eq:E-tot-sum}
\end{equation}

The electrostatic energy of the cluster ionic core is equal to
\begin{equation}
E_{core} = \frac{1}{2} \int_V \rho({\bf r}) U_{core}({\bf r}) {\rm d}{\bf r}.
\label{eq:E-core}
\end{equation}

In the Hartree-Fock approximation, the
electronic energy $E_{el}$ is given by
the general expression \cite{Lindgren}:
\begin{eqnarray}
E^{HF}_{el} & = &
\sum_{a} < a \mid - \Delta/2 + U_{core} \mid a> + \nonumber \\
&& \frac{1}{2} \sum_{abk} q_a q_b
\left[ c(abk) F^{k}(a,b) + d(abk) G^{k}(a,b) \right],
\label{eq:E-el}
\end{eqnarray}
where $a$ and $b$ run over all shells. The values
$F^{k}(a,b)$ and $G^{k}(a,b)$ in the Eq. (\ref{eq:E-el})
are the Coulomb and exchange
Slater integrals, $q_a$ and $q_b$ are the occupation numbers
for orbitals $a$ and $b$, respectively. The Hatree-Fock coefficients 
$c(abk)$ and $d(abk)$ for the Coulomb and exchange energy contributions
depend on the occupation numbers (see for details \cite{Lindgren}).

In the framework of LDA the electronic energy of the system 
is given by \cite{Hohenberg-Kohn,Kohn-Sham}: 
\begin{eqnarray}
E^{LDA}_{el} & = &
\sum_{a} < a \mid - \Delta/2 + U_{core} \mid a> + \nonumber \\
&& \frac{1}{2} \int
\frac{\rho_{el}({\bf r}) \rho_{el}({\bf r'})}{{\bf |r-r'|}}
{\rm d}{\bf r} {\rm d}{\bf r'} +
\int \rho_{el}({\bf r}) \epsilon_{xc}(\rho_{el}({\bf r})) {\rm d}{\bf r},
\label{eq:E-el-LDA}
\end{eqnarray}

\noindent where the latter term represents the exchange-correlation energy.

\section{Numerical results}

Let us present and discuss the results of calculations performed
in the model described above. We start our consideration with
the analysis of electronic configurations
and occupation numbers alterations during the
fission processes of the doubly charged
sodium clusters $Na_{10}^{2+}$ and $Na_{18}^{2+}$,
which we perform using  the two overlapping sphere
parametrization model. The second part of our discussion is devoted to
the cluster energetics
and formation of fission barriers, for the fission channels considered.
We compare the 
results obtained for the
two overlapping spheres and the two overlapping
spheroids parametrization,
as well as the variable-necking-in type of the shape parametrization and
demonstrate the crucial importance of the cluster
shape deformations in the fission process.

\begin{figure}
\begin{center}
\includegraphics[scale=0.36]{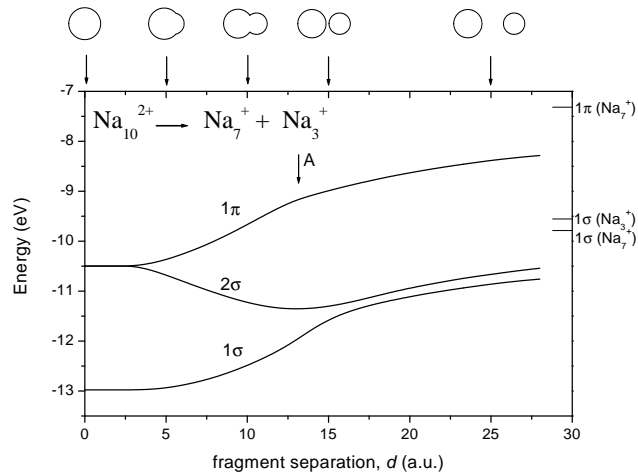}
\end{center}
\caption{Hartree-Fock single electron energy levels for the
asymmetric fission channel
$Na_{10}^{2+} \rightarrow Na_{7}^{+} + Na_{3}^{+}$
as a function of fragments separation distance $d$, when both the parent and
daughter clusters are spherical ($\delta_{0}=\delta_{1}=\delta_{2}=0$).
The evolution of the cluster shape during the fission process is
shown on top of the figure.
Horizontal lines on the right hand side of the figure mark
the Hartree-Fock energy levels for the free daughter fragments.}
\label{fig:levels10asym}
\end{figure}

\begin{figure}
\begin{center}
\includegraphics[scale=0.36]{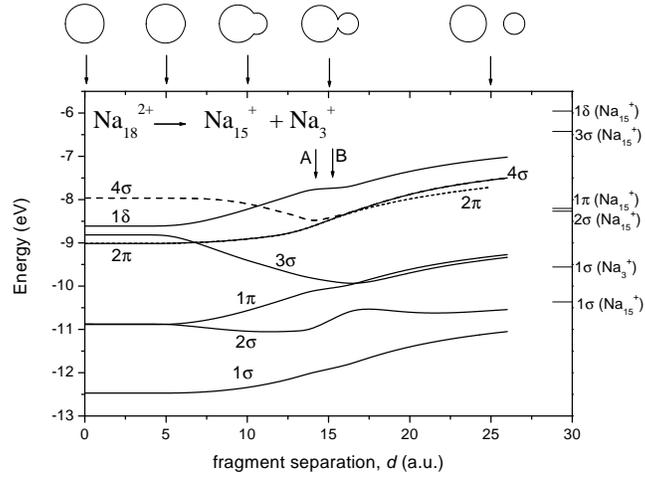}
\end{center}
\caption{The same as figure \ref{fig:levels10asym}
but for asymmetric fission channel
$Na_{18}^{2+} \rightarrow Na_{15}^{+} +Na_{3}^{+}$.
Lowest  unoccupied states are shown by dashed ($4\sigma$) and short dashed
($1\delta$) lines.}
\label{fig:levels18asym}
\end{figure}

\begin{figure}
\begin{center}
\includegraphics[scale=0.36]{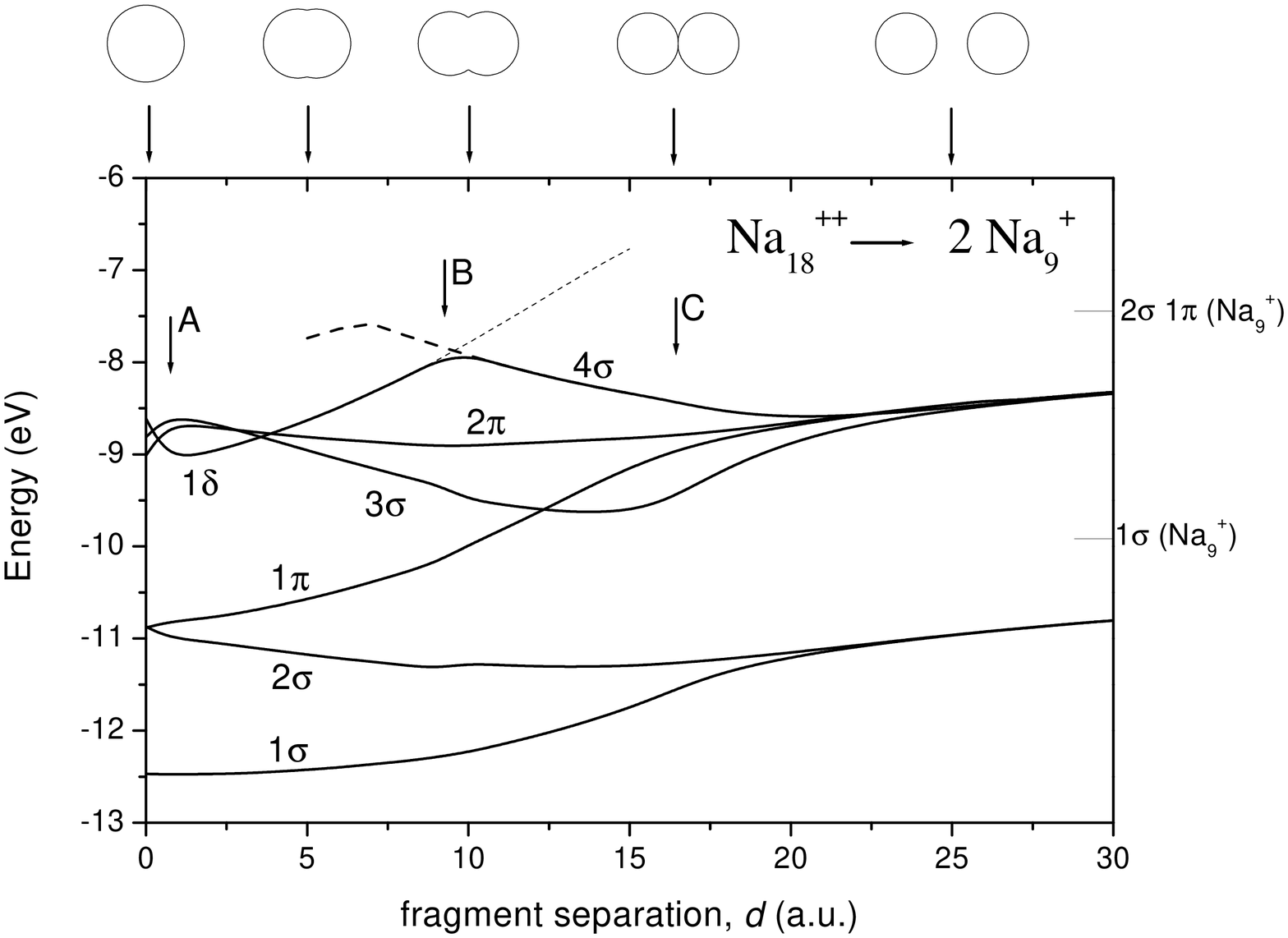}
\end{center}
\caption{The same as figure \ref{fig:levels10asym}
but for 
symmetric fission channel $Na_{18}^{2+} \rightarrow 2 Na_{9}^{+}$.
Lowest  unoccupied states are shown by dashed ($4\sigma$) and short dashed
($2\pi$) lines.}
\label{fig:levels18sym}
\end{figure}

Figures \ref{fig:levels10asym}--\ref{fig:levels18sym}
show the Hartree-Fock single-electron energies
$\varepsilon_i$ as a function of the fragment separation distance $d$
for the following asymmetric 
$Na_{10}^{2+} \rightarrow Na_{7}^{+} + Na_{3}^{+}$,
$Na_{18}^{2+} \rightarrow Na_{15}^{+} + Na_{3}^{+}$
and symmetric
$Na_{18}^{2+} \rightarrow 2 Na_{9}^{+}$
channels respectively, when
both parent and daughter clusters are assumed to be
spherical, and, hence, for any $d$ the deformation parameters are equal to
zero $\delta_{0}=\delta_{1}=\delta_{2}=0$.  The snapshots of the cluster
shape evolution during the fission are shown on top of each figure.

The initial electronic configuration for the parent $Na_{10}^{2+}$ cluster is
$1\sigma^2 2\sigma^2 1\pi^4$, where $2\sigma 1\pi$ level is degenerated,
due to spherical symmetry of the system with closed electronic shells.  
With increasing fragments separation distance $d$,
the spherical symmetry of the parent cluster breaks down,
that results in splitting of the single electron energy levels according to
projection of the angular momentum along the $z$-axis.
A similar splitting of the energy levels has been
studied for deformed nuclei (see e.g. \cite{Migdal}) and clusters
\cite{Lyalin00,Lyalin01a}. For high enough separation distances
beyond the scission point, indicated by vertical arrow A
in figure \ref{fig:levels10asym}, 
energy levels approach to their limiting values
marked by horizontal lines on the right hand side of the figure,
being  the electron energy levels of the free fission fragments
$Na_{7}^{+}$ ($1\sigma^2 1\pi^4$) and $Na_{3}^{+}$ ($1\sigma^2$).
 
The electronic configuration and the orbital occupation numbers
for $Na_{18}^{2+}$ cluster exhibits
several alterations during the fission process, as it is shown
in figures \ref{fig:levels18asym} and \ref{fig:levels18sym}.
This happens due to the fact that different electronic
configurations minimize the total energy of the cluster at
different separation distances.
The parent 
cluster $Na_{18}^{2+}$ has the initial electronic configuration
$1\sigma^2 2\sigma^2 1\pi^4 2\pi^2 3\sigma^2  1\delta^4 $.
Following to the asymmetric channel of fission
$Na_{18}^{2+} \rightarrow Na_{15}^{+} + Na_{3}^{+}$
(see figure \ref{fig:levels18asym})
the new configuration
$1\sigma^2 2\sigma^2 1\pi^4 3\sigma^2 4\sigma^2 1\delta^4 $
becomes preferable, when two electrons transfer from the
half-filled $2\pi$ state
to initially unoccupied $4\sigma$ state.
This happens for the separation distance $d \ge 14.2$ a.u.
(marked in figure \ref{fig:levels18asym} by the solid vertical arrow A)
before the scission point (vertical arrow B) $d=15.3$ a.u.
The order of the energy levels manifests several re-distributions
during the fission
process, and finally for high enough separation distance ($d>25$ a.u.), its
energy levels correspond to the free fission fragments
$Na_{15}^{+}$ ($1\sigma^2 2\sigma^2 1\pi^4 3\sigma^2 1\delta^4$ )
and $Na_{3}^{+}$ ($1\sigma^2$).

For the symmetric fission channel
$Na_{18}^{2+} \rightarrow 2 Na_{9}^{+}$
(figure \ref{fig:levels18sym})
the energy levels show even more complicated behaviour. 
Thus, for the separation distance $d \ge 0.7$ a.u.
(marked in figure \ref{fig:levels18sym} by solid vertical arrow A)
the new intermediate  configuration
$1\sigma^2 2\sigma^2 1\pi^4 1\delta^2 2\pi^4 3\sigma^2 $
becomes preferable, when two electrons transfer from the occupied $1\delta$
to the half-filled $2\pi$ state.
Further increasing of the separation distance,
leads to the transition of the
two residual $1\delta$ electrons to the initially unoccupied $4\sigma$
state for $d \ge 9.25$ a.u.  (marked in figure \ref{fig:levels18sym}
by the solid vertical arrow B) forming the final
electronic configuration
$1\sigma^2 2\sigma^2 1\pi^4 3\sigma^2 2\pi^4 4\sigma^2$
before the scission point (vertical arrow C) $d=16.64$ a.u.
The order of the energy levels manifests several re-distributions
during the fission
process, and finally for high enough separation distance ($d>25$ a.u.), its
ordering corresponds to $1\sigma 2\sigma 1\pi$ which determined by the
magic spherical $Na_{9}^{+}$ products.
They are doubly degenerated as compared to the
initial configuration, since there are two $Na_9^{+}$ fragments in the
system. Such a behaviour of the HF levels as a function of $d$ is quite
similar to those following from the ATCOSM simulations \cite{Yannouleas95a}.



\begin{figure}
\begin{center}
\includegraphics[scale=0.36]{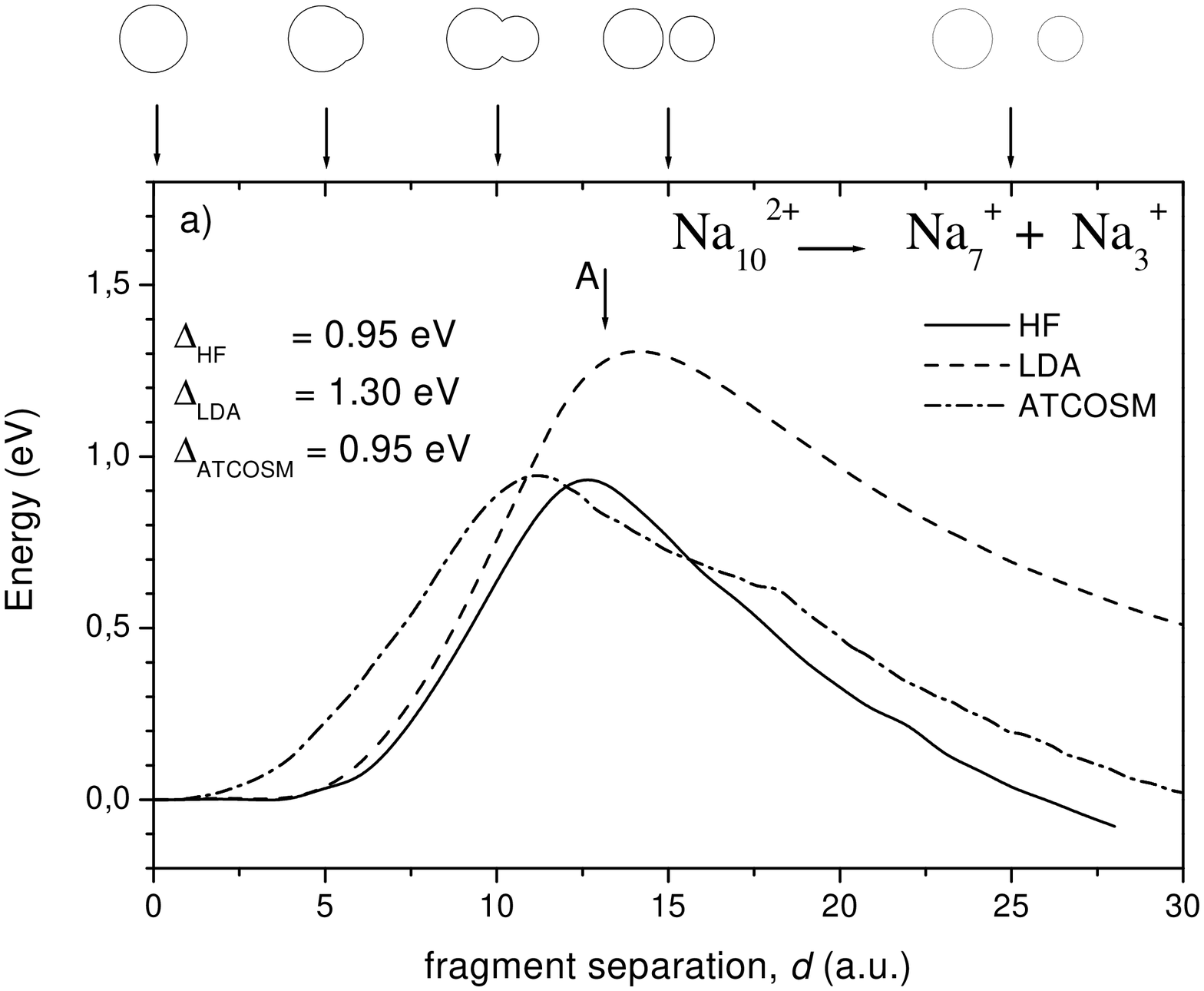}
\includegraphics[scale=0.36]{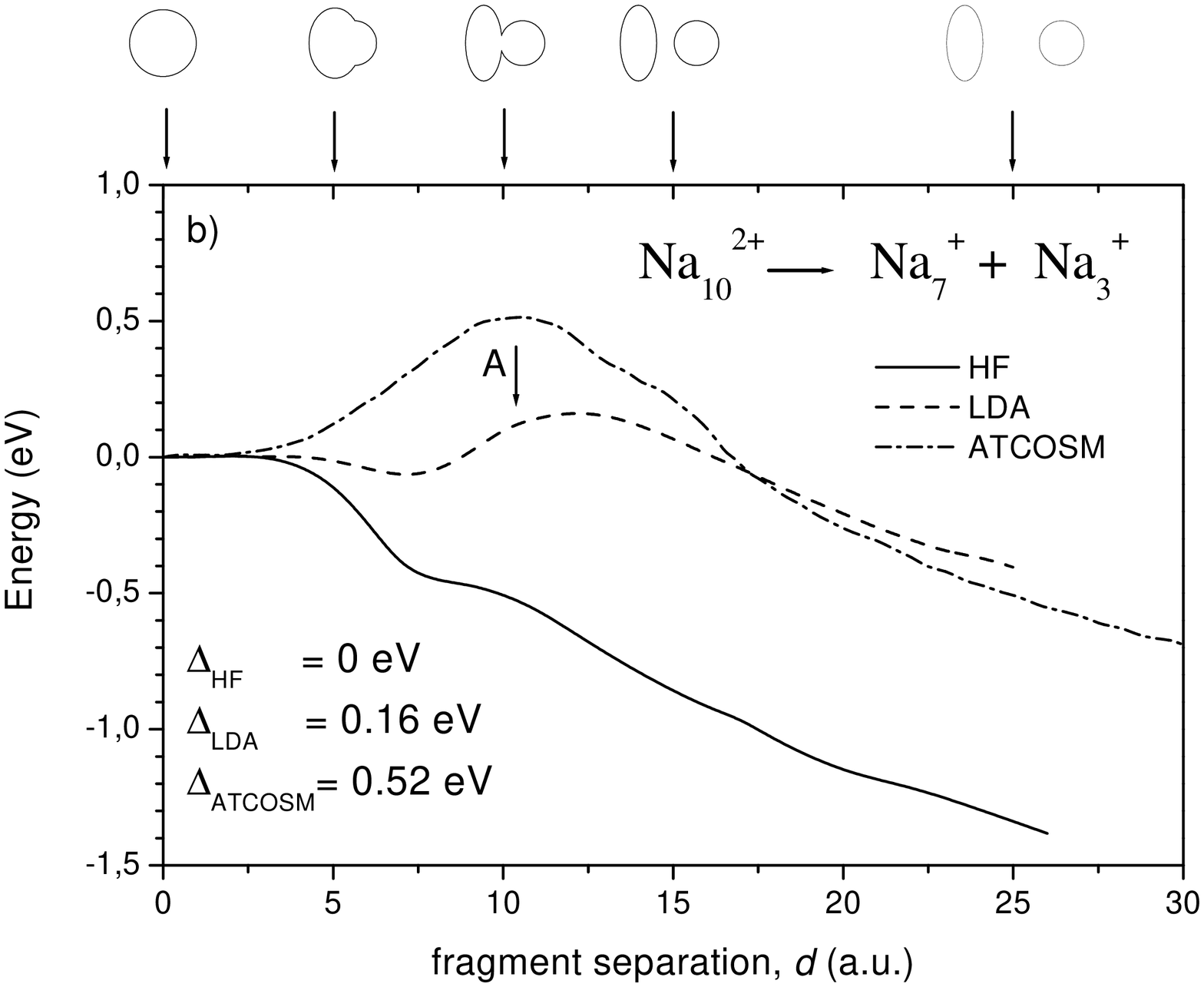}
\end{center}
\caption{Fission barriers in the two-center deformed jellium
Hartree-Fock (solid lines) and LDA (dashed lines) approaches
calculated in this work for the asymmetric channel
$Na_{10}^{2+} \rightarrow Na_{7}^{+} + Na_{3}^{+}$
as a function of fragments separation distance $d$.
In (a),  both parent and
daughter clusters are spherical, $\delta_{0}=\delta_{1}=\delta_{2}=0$.
In (b), deformations of the parent and daughter clusters are taken
into account.
The zero of energy put at $d=0$.
The evolution of cluster shape during the fission process is
shown on top of figures (a) and (b) for both models. We compare our
results with those derived in ATCOSM (dash-dotted line) \cite{Yannouleas95a}.}
\label{fig:bar-10asym}
\end{figure}

Figure \ref{fig:bar-10asym} presents fission barriers
for the asymmetric channel 
$Na_{10}^{2+} \rightarrow Na_{7}^{+} + Na_{3}^{+}$ 
as a function of the fragments separation distance $d$.
In order to perform the accurate comparison of fission barriers
derived in the 
two-center deformed jellium HF and LDA models
with the ATCOSM results \cite{Yannouleas95a},
we have used two different type of shape parametrization.
Thus, the upper part in figure \ref{fig:bar-10asym}
shows the barriers for fission of a spherical
parent cluster into two spherical daughter fragments 
(i.e. $\delta_{0}=\delta_{1}=\delta_{2}=0$ in our model).
This type of parametrization, known as the two-intersected spheres
parametrization,
was used in many works \cite{Saito87,Saito88,Knospe93}.
The low part in figure \ref{fig:bar-10asym} shows fission barriers
derived on the basis of parametrization accounting for an 
independent deformation
of parent and daughter clusters
in order to minimize the total energy of the system
for any distance $d$.  The evolution of cluster shape during
the fission process is shown
for the HF and LDA models
on tops of the
corresponding figures. Note, that 
the variable-necking-in parametrization has been used
in the ATCOSM calculation
\cite{Yannouleas95a}.

Solid lines in figure \ref{fig:bar-10asym}  are the result of the
two-center jellium HF model, while dashed curves have been calculated
in LDA.
Dash-dotted lines show the ATCOSM barriers calculated in 
\cite{Yannouleas95a}. The zero of energy put at $d=0$.

Figure \ref{fig:bar-10asym} (a) demonstrates a good agreement of the
HF and ATCOSM 
fission barriers heights $\Delta_{HF}=\Delta_{ATCOSM}=0.95$ eV.
The LDA value for the fission barrier,
$\Delta_{LDA}=1.30$ eV, slightly exceeds the HF and ATCOSM ones.
In HF and LDA schemes the fission barrier maximum is
located just behind the scission point (marked by vertical arrow A).

The two-intersected spheres parametrization is not adequate for 
the description of the process
$Na_{10}^{2+} \rightarrow Na_{7}^{+} + Na_{3}^{+}$,
because $Na_{7}^{+}$ daughter fragment has an open-shell electronic
structure, and therefore its shape is not spherical
because it undergoes Jahn-Teller
distortion \cite{Frauendorf96}. The deformed jellium
Hartree-Fock and LDA calculations show that the ground state
of the $Na_{7}^{+}$ cluster has the oblate shape
with the deformation parameter $\delta_1=-0.68$. This value is
in a good agreement with other theoretical estimates
\cite{Lyalin00}.
The oblate shape deformation of the daughter cluster
reduces the final total energy of the system $E_{tot}$
by $-1.32$ eV for the HF, and by $-1.05$ eV for the LDA model
calculations, in comparison with the spherical case.


Part (b), figure \ref{fig:bar-10asym} shows 
fission barriers for the asymmetric channel
$Na_{10}^{2+} \rightarrow Na_{7}^{+} + Na_{3}^{+}$, when spheroidal
deformations of the parent and daughter clusters
are taken into account. We have minimized
the total energy of the system over 
the parent and daughter fragments deformations 
with the aim to find the fission pathway corresponding
to the minimum of the fission barrier.
We have also used the assumption of continuous shape deformation
during the fission process.
The deformation of the cluster fragments changes drastically the fission
energetics in comparison with what follows from the two-intersected spheres
model.
In the framework of the two-center deformed jellium Hartree-Fock
approximation, the parent cluster $Na_{10}^{2+}$ becomes unstable towards
the asymmetric channel
$Na_{10}^{2+} \rightarrow Na_{7}^{+} + Na_{3}^{+}$.

The LDA simulations with deformation of the daughter fission fragments
take into account the decrease of the total energy of the system (see
figure \ref{fig:bar-10asym}b).
In particular, this results in the appearance of
the local minimum in the energy curve at $d=7.2$ a.u.,  
corresponding to the formation of the
super deformed asymmetric prolate shape of the parent $Na_{10}^{2+}$
cluster before the scission point A.
The latter is located at $d=10.4$ a.u. The allowance for  
deformation of the parent cluster and the fission fragments reduces the
LDA fission barrier up to the value $\Delta_{LDA}=0.16$ eV, which
is in rather poor agreement with the result of ATCOSM 
$\Delta_{ATCOSM}=0.52$ eV.

It is necessary to note that 
the shape of the cluster fragments in ATCOSM has been parametrized by two
spheroids of revolution connected by a smooth neck \cite{Yannouleas95a}.
Our calculations show that the oblate shape of $Na_{7}^{+}$ fragment
is formed at the initial stage of fission process, for separation distances
before the scission point. Moreover, in the vicinity of the scission point,
where the interaction between the two daughter fragments
$Na_{7}^{+}$ and $Na_{3}^{+}$
is strong, the oblate $Na_{7}^{+}$ fragment is even more deformed
than a free one. This means that it is
more favourable for two fragments to split at shorter distances
rather than
to be connected by a smooth neck, making the system more prolate.
This fact explains why using of 
necking type of shape parametrization, leads to the higher barrier
as compare to our parametrization.

\begin{figure}
\begin{center}
\includegraphics[scale=0.36]{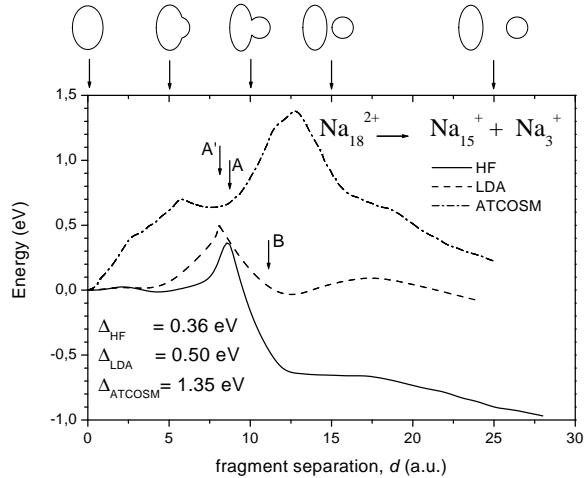}
\end{center}
\caption{The same as figure \ref{fig:bar-10asym}, but for
the asymmetric channel
$Na_{18}^{2+} \rightarrow Na_{15}^{+} + Na_{3}^{+}$}
\label{fig:bar-18asym}
\end{figure}

Comparison of the asymmetric
$Na_{18}^{2+} \rightarrow Na_{15}^{+} + Na_{3}^{+}$
and symmetric $Na_{18}^{2+} \rightarrow 2 Na_{9}^{+}$ fission channels
of the parent $Na_{18}^{2+}$ cluster is a subject of 
particular interest,
because there have to be a 
competition between these two channels involving magic cluster
ions $Na_{3}^{+}$ and $Na_{9}^{+}$.
In \cite{Barnett91,Brechignac91,Yannouleas95} it was noticed that 
namely in fission of the $Na_{18}^{2+}$ cluster 
a magic fragment other than $Na_{3}^{+}$ becomes
the favoured channel.
Figure \ref{fig:bar-18asym} shows fission barriers
for the asymmetric channel 
$Na_{18}^{2+} \rightarrow Na_{15}^{+} + Na_{3}^{+}$. 
We have started from the initial configuration corresponding to oblate
shape of the parent $Na_{18}^{2+}$ cluster with the deformation parameter
$\delta_0=-0.35$. The oblate deformation reduces the
total energy of the cluster $Na_{18}^{2+}$ by
$-0.58$ eV for the HF and by $-0.48$ eV for the LDA model
in comparison with the spherical case.
We have minimized the total energy of the system over the deformation
parameters of the parent and daughter fragments during the fission
process for any separation distance $d$.
The evolution of the fragment shapes
is shown on top of the figure for both HF and LDA models.
The daughter fragment $Na_{15}^{+}$ has an oblate shape with deformation
parameter $\delta_1=-0.6$, while  $Na_{3}^{+}$ is spherical, i.e.
$\delta_2=0$.

The total energy as a function of
the fragment separation distance has a maximum
(marked by vertical arrow A for HF, and A$^{'}$
for LDA), arising due to the
alteration of the electronic configuration
$1\sigma^2 2\sigma^2 1\pi^4 2\pi^2 3\sigma^2  1\delta^4 $
$\rightarrow$
$1\sigma^2 2\sigma^2 1\pi^4 3\sigma^2 4\sigma^2 1\delta^4 $.
These maxima on the energy curves define the fission barrier hights,
being equal to $\Delta_{HF}=0.36$ eV for Hartree-Fock  and
$\Delta_{LDA}=0.50$ eV for LDA.
It is interesting to notice that the LDA total energy curve
has a pronounced minimum at $d=12.5$ a.u., located beyond the scission point
$d=11.1$ a.u. We have marked the scission point by vertical arrow B,
both for HF and LDA.
This minimum means that a quasistable state of
the supermolecule $Na_{15}^{+} + Na_{3}^{+}$
can be created during the fission process.
However, the appearance of the minimum and thus the stability of the super
molecule is rather sensitive to the model chosen for
the description of exchange and correlation 
inter-electron interaction. This is already clear from the fact that
such a minimum does not appear in the HF simulations
(see figure \ref{fig:bar-18asym}).   

The ATCOSM model calculation gives the value of the fission barrier  
$\Delta_{ATCOSM}=1.35$ eV, which is inconsistent with the HF and LDA
results derived in our model. Such a difference can be explained as a result
of variable-necking-in type of the shape parametrization, which has
been used in
\cite{Yannouleas95a}.  In the case of asymmetric 
$Na_{18}^{2+} \rightarrow Na_{15}^{+} + Na_{3}^{+}$  channel,
the parent as well as one of the daughter fragments have an 
oblate shape, therefore shape parametrization model with prolate-like
additional neck is not natural, and results in increasing the fission barrier.


\begin{figure}
\begin{center}
\includegraphics[scale=0.36]{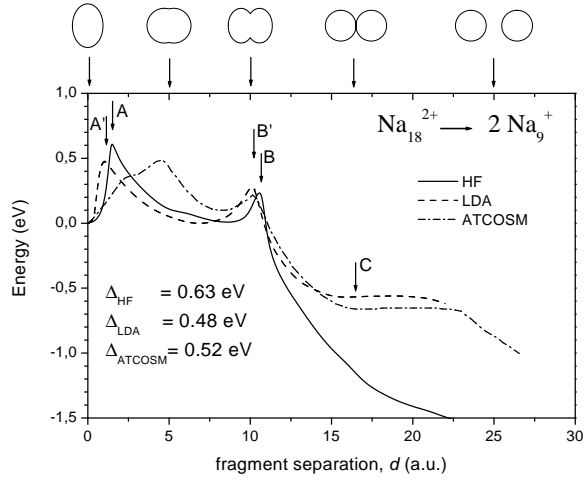}
\end{center}
\caption{The same as figure \ref{fig:bar-10asym}, but for
the symmetric channel
$Na_{18}^{2+} \rightarrow 2 Na_{9}^{+}$}
\label{fig:bar-18sym}
\end{figure}

Figure \ref{fig:bar-18sym} shows the  dependence of 
total energy $E_{tot}$ on separation distance $d$
for the symmetric channel
$Na_{18}^{2+} \rightarrow 2 Na_{9}^{+}$.
The parent cluster changes its shape from
oblate to prolate one on the initial stage of the fission process
($d\approx 1$ a.u.). This transition is accompanied by the first
re-arrangement
of the electronic configuration (marked by vertical arrow A for HF
and A$^{'}$ for LDA).  This process has the barrier
$\Delta_{HF}=0.63$ eV and $\Delta_{LDA}=0.48$ eV.
On the next stage of the reaction the prolate deformation develops
resulting in the highly deformed cluster shape,
as it is shown
on top of figure \ref{fig:bar-18sym}.
At the distance $d\approx 11$  (marked by vertical arrow B for HF,
and B$^{'}$ for LDA)
the  electronic configuration reaches its final form being the same as in
the spherical $Na_{9}^{+}$ products. In this case, the variable-necking-in
type of fragments shape parametrization used in ATCOSM 
does not brake the symmetry of the fragments
and therefore the agreement between the two overlapped spheroids HF or LDA
models and ATCOSM variable-necking-in approach is much better than in the case
of asymmetric fission channel.    
Indeed, the total fission barrier for the symmetric channel
$Na_{18}^{2+} \rightarrow 2 Na_{9}^{+}$
is equal to $\Delta_{HF}=0.63$ eV and 
$\Delta_{LDA}=0.48$ eV
in the two-center jellium Hartree-Fock and LDA models respectively.
These  values are in a good agreement with the ATCOSM result
$\Delta_{ATCOSM}=0.52$ eV \cite{Yannouleas95a}.


In table \ref{tab:barrier} we have summarized the results of
the HF and LDA barrier heights calculations for the considered 
fission channels and compared them with the results of ATCOSM
\cite{Yannouleas95a} and MD simulations
\cite{Montag95a,Guet01,Barnett91}.
 
\begin{table*}
\label{tab:barriers}
\caption{\label{tab:barrier} Summary of the fission barrier heights in (eV)
calculated in this work and their comparison with the results of other
approaches.}
\begin{ruledtabular}
\begin{tabular}{lcccccc}
Channel & HF & LDA & ATCOSM \cite{Yannouleas95a} &
MD \cite{Montag95a} & MD \cite{Guet01} & MD \cite{Barnett91}\\
$Na_{10}^{2+} \rightarrow Na_{7}^{+} + Na_{3}^{+}$
& 0 & 0.16 & 0.52 & 0.67 & 0.50& 0.71\\
$Na_{18}^{2+} \rightarrow Na_{15}^{+} + Na_{3}^{+}$
& 0.36 & 0.50&1.35 & 0.50 & -- & -- \\
$Na_{18}^{2+} \rightarrow 2 Na_{9}^{+}$
& 0.63 &0.48 &0.52 &0.52 & -- & --\\
\end{tabular}
\end{ruledtabular}
\end{table*}

The height of the fission barrier for $Na_{10}^{2+}$ cluster
in the two-center deformed jellium LDA model
is $0.51$ eV lower than its value following from the MD simulations
\cite{Montag95a}.
Molecular dynamics simulations performed in \cite{Montag95a}
also were based on the density-functional theory,
but included full ionic structure of the cluster.
Since both methods apply the same 
form of the density functional \cite{Gunnarsson},
the discrepancy in the fission barrier heights
can be attributed to the manifestation
of the influence of the cluster ionic structure in the fission process.  
One can expect that the influence
of the detailed ionic structure has to decrease with the growth cluster size,
making the jellium model approach more and more accurate.
However, we also want to notice here that different schemes of MD simulations
\cite{Montag95a,Guet01,Barnett91} lead to somewhat different fission barrier
heights (see Table \ref{tab:barriers}).

For the $Na_{18}^{2+}$ cluster, we report
a very good agreement of the heights of fission barriers
derived in the jellium LDA model
and MD \cite{Montag95a}.  
It is interesting to note that MD simulations predict
the asymmetric channel $Na_{18}^{2+} \rightarrow Na_{15}^{+} + Na_{3}^{+}$
to be a little more favourable, however the jellium LDA model says in favour of
the symmetric path $Na_{18}^{2+} \rightarrow 2 Na_{9}^{+}$.
Although, in both models the difference in the heights of symmetric
and asymmetric fission barriers is rather small, being equal to $0.02$ eV.

Finally, let us compare the results obtained in the 
HF and LDA models. The Hartree-Fock model takes
into account the exchange inter-electron interaction the most exactly,
but it neglects many-electron correlations, playing quite a significant role
in metal cluster energetics. Thus, LDA calculations with the 
Gunnarsson-Lundqvist's exchange-correlation potential
\cite{Gunnarsson} lower the total energy $E_{tot}$
of the parent $Na_{10}^{2+}$  and  $Na_{18}^{2+}$ clusters
on $-5.58$ eV and $-12.03$ eV respectively as
compared to the energies calculated in the Hartree-Fock approximation.
Such a difference between the total energies of the HF and LDA jellium
clusters is a result of the manifestation of many-electron correlation
interaction. The many-electron correlation interaction reduces 
at distances $d$ beyond the scission point, when the parent cluster 
splits on the two fragments, due to decrease of the number of interacting
electrons in the system.

\begin{figure}
\begin{center}
\includegraphics[scale=0.36]{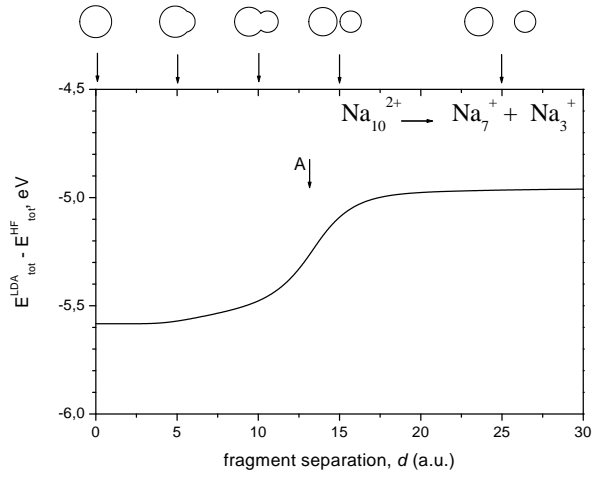}
\end{center}
\caption{Difference between the LDA and HF total energies
as a function of fragments separation distance
for the fission channel $Na_{10}^{2+} \rightarrow Na_{7}^{+} + Na_{3}^{+}$.
The evolution of the cluster shape during the fission process is shown on
top of the figure. Vertical arrow A corresponds to the scission point.}
\label{fig:cor-10sphere}
\end{figure}

Figure \ref{fig:cor-10sphere} shows the 
difference between the LDA and HF total cluster energies
$E^{LDA}_{tot} - E^{HF}_{tot}$ as a function of fragments separation distance
for the fission channel $Na_{10}^{2+} \rightarrow Na_{7}^{+} + Na_{3}^{+}$.
For illustrative purposes we use here the two overlapping sphere 
parametrization to dismiss the effect of the clusters fragment
shape alteration along the fission pathway, which can be different
in the HF or LDA schemes of calculation.

It is seen from figure \ref{fig:cor-10sphere} that the
difference between the LDA and HF total energies 
$E^{LDA}_{tot} - E^{HF}_{tot}$ 
does not change significantly for separation
distances below the scission point, in spite of the strong deformation
of the parent cluster. In the narrow region of $d$, nearby the scission
point the value $E^{LDA}_{tot} - E^{HF}_{tot}$ increases
and it becomes constant again above the scission point.
Such a behavior is in a great deal similar,
at least qualitatively, to the all considered fission channels. 
It remains valid even if deformations of the parent and daughter fragments
are taken into account.
This fact has a simple physical explanation. The many-electron correlations
reduce the total energy.
During the fission process
the many-electron correlation interaction
between electrons belonging the two different cluster fragments vanishes,
which results in the increase of the total energy of the system.  

In spite of the fact that many-electron correlations reduce significantly
the total energy of the $Na_{10}^{2+}$ and $Na_{18}^{2+}$
jellium clusters in comparison with the HF values,
the difference in heights of the fission barriers
obtained in the HF and LDA models is only about $0.15$ eV.
From the analysis carried out above one can conclude that
many-electron correlations do not influence significantly
on the height of the fission barrier if the latter arises well
below before the scission point.
In the cases when the barrier is created nearby or above
the scission point, accounting for 
many-electron correlations becomes essential.   
For example, many-electron correlations play the crucial role for the 
$Na_{10}^{2+} \rightarrow Na_{7}^{+} + Na_{3}^{+}$ fission channel.
In this case the HF model predicts qualitatively wrong barrierless
scenario of the fission.
In the contrary, the agreement between the HF and LDA models is
much better
for both symmetric and asymmetric fission channels of the $Na_{18}^{2+}$
cluster. Note also that the HF model,
predicts the asymmetric fission channel
$Na_{18}^{2+} \rightarrow Na_{15}^{+} + Na_{3}^{+}$
to be more favourable in comparison with the symmetric one.

\section{Summary}

We have developed the open-shell two-center deformed jellium
Hartree-Fock and LDA method for the description of metal
clusters fission process.
The proposed two overlapping sheroids shape parametrization
allows one to consider independently a wide variety of shape
deformations of parent and daughter clusters, and
to investigate the role of deformation effects
on the  cluster fission process. The proposed type of
shape parametrization is preferable as compared to 
variable-necking-in one, in particular when the fission fragments have an
oblate shape.
The role of many-electron correlation effects in metal clusters fission
is analysed. The described Hartree-Fock model forms the basis for
further systematic
development of the more advanced {\it ab initio} many-body theories for the
process of metal clusters fission.

\begin{acknowledgments}
The authors acknowledge financial support from
the Volkswagen Foundation, the Alexander von Humboldt Foundation,
INTAS, DFG, DAAD, and the Royal Society of London.
\end{acknowledgments}

\end{document}